\documentclass[aps,prb,twocolumn,showpacs,amsmath,amssymb,superscriptaddress]{revtex4-1}
\usepackage{graphicx}% Include figure files
\usepackage{dcolumn}% Align table columns on decimal point
\usepackage{epstopdf}
\usepackage{bm}% bold math
\usepackage{color}

\begin{document}
\title{Adsorption of cobalt on graphene: Electron correlation effects \\ from a quantum chemical perspective}

\author{A.~N. Rudenko}
\email[]{rudenko@tu-harburg.de}
\author{F.~J. Keil}
\affiliation{Institute of Chemical Reaction Engineering, Hamburg University of Technology, Eissendorfer Strasse 38, D-21073 Hamburg, Germany}
\author{M.~I. Katsnelson}
\affiliation{Institute for Molecules and Materials, Radboud University Nijmegen, Heijendaalseweg 135, 6525 AJ Nijmegen, The Netherlands}
\author{A.~I. Lichtenstein}
\affiliation{Institute of Theoretical Physics, University of Hamburg,
Jungiusstrasse 9, D-20355 Hamburg, Germany}
\date{\today}

\begin{abstract}
In this work, we investigate the adsorption of a single cobalt atom (Co) on graphene by means of the complete active space self-consistent field approach, 
additionally corrected by the second-order perturbation theory. The local structure of graphene is modeled by a planar hydrocarbon cluster (C$_{24}$H$_{12}$). 
Systematic treatment of the electron correlations and the possibility to study excited states allow us to reproduce 
the potential energy curves for different electronic configurations of Co.
We find that upon approaching the surface, the ground-state configuration of Co undergoes several transitions, giving rise to two
stable states. The first corresponds to the physisorption of the adatom in the high-spin $3d^74s^2$ ($S=3/2$) configuration, while the second results from the chemical bonding formed by strong
orbital hybridization, leading to the low-spin $3d^9$ ($S=1/2$) state. Due to the instability of the $3d^9$ configuration, the adsorption energy of Co is small in both cases and does not exceed 0.35 eV. 
We analyze the obtained results in terms of a simple model Hamiltonian that involves Coulomb repulsion ($U$) and exchange coupling ($J$) parameters for the 3$d$ shell of Co, which we estimate from 
first-principles calculations. We show that while the exchange interaction remains constant upon adsorption ($\simeq1.1$~eV), the Coulomb repulsion significantly reduces for decreasing distances (from 5.3 to 2.6$\pm$0.2 eV).
The screening of $U$ favors higher occupations of the 3$d$ shell and thus is largely responsible for the interconfigurational transitions of Co. 
Finally, we discuss the limitations of the approaches that are based on density functional theory with respect to transition metal atoms on graphene, and 
we conclude that a proper account of the electron correlations is crucial for the description of adsorption in such systems.
\end{abstract}

\pacs{31.15.A-, 31.15.ae, 34.35.+a, 81.05.ue}
\maketitle

\section{Introduction}

Graphene is the focus of many research activities due to its remarkable electronic properties.\cite{Graphene-RMP,Katsnelson-Book} 
The magnetic properties of graphene also attract considerable attention because of their potential use in spintronics.\cite{Tombros}
The intrinsic magnetism of graphene, being theoretically predicted for localized defects or edge-states,\cite{Yazyev} represents a challenging
problem for experimental studies due to the enhanced chemical reactivity of magnetic centers that quenches magnetism in the presence of minor contamination. 
Furthermore, no magnetic ordering has been observed in defective graphene even at liquid helium temperatures,\cite{Sepioni,Nair2012} which suggests that 
this kind of magnetism can hardly be employed in practical applications.
One of the natural ways to induce a magnetic moment in a closed-shell compound is to include magnetic impurities, such as transition metal (TM) atoms, to the system. 
Apart from being magnetically stable, deposition of TM adatoms on the graphene surface may give rise to a variety of many-body effects 
(such as, e.g., the Kondo effect),\cite{APS,Cornaglia,Wehling2010} known mainly for TM adatoms supported on metallic substrates.\cite{J-Li,Madhavan}

In contrast to molecular closed-shell adsorbates,\cite{Rudenko2010} light monovalent impurities,\cite{Wehling2009} or even some 
insulating substrates,\cite{Rudenko2011a,Rudenko2011b} a theoretical 
investigation of TM adatoms on graphene is more challenging due to the presence of strong electron correlations.
Commonly used \emph{ab initio} concepts, such as density functional theory (DFT) or single-determinant Hartree-Fock (HF) \mbox{-based}
approaches, are often unreliable even in terms of a qualitative description of the TM complexes.
DFT results may vary depending on the parametrization of the exchange-correlation functional, whereas the applicability of the HF-based methods is
questionable for systems with near-degeneracies. 
The latter is attributed to the inherent limitation of single-reference approaches, such as the inability to treat correlation effects
arising from the mixture of low-lying electronic configurations.
Although the model Hamiltonian approaches are appealing and widely used to investigate strongly correlated systems, their results often
depend on a number of adjustable parameters, which prevent these approaches from being a stand-alone tool for studying real materials.

Among the other first-row transition metals, cobalt holds a special place due to
the remarkable magnetic properties of its compounds related to high Curie temperatures and large magnetic anisotropies.
The experimentally observed Kondo effect in the system of Co on graphene,\cite{APS} followed by considerable efforts in the description of this 
phenomenon,\cite{Wehling2010,Jacob,Vojta} make Co on graphene one of the most attractive candidates for theoretical studies among the other TM/graphene 
systems.

While there are only a few experimental works,\cite{APS,Brar} there have been numerous theoretical DFT-based studies that were focused on Co on pristine 
graphene.\cite{Wehling2010,Mao,Johll,Cao,Yazyev2010,Valencia,Chan2011a,Liu,Chan2011b,Ding,Wehling2011,Lima,Saffarzadeh}
With respect to chemical bonding, DFT studies predict strong covalent interaction arising from the
hybridization between the cobalt 3$d$- and graphene $\pi$-orbitals, resulting in a low-spin adatom configuration ($S=1/2$). The hybridization leads to relatively 
large binding energies of $\sim$1--2 eV, depending on the initial electronic configuration of the Co atom, and also on the particular parametrization of the 
DFT exchange-correlation functional. Large binding energies obtained within the DFT generally imply high migration barriers and, therefore, thermodynamic stability of Co atoms 
on graphene. However, this implication seems to be inconsistent with the existing experimental evidence of fast surface diffusion of transition metal adatoms on graphene, 
as concluded from transmission electron microscopy experiments.\cite{Anton,Gan,Manzo}
This observation suggests a weak bonding, and that there may be another mechanism behind the Co--graphene interaction.

Not many attempts have been made thus far to go beyond the standard local [local density approximation (LDA)] or semilocal 
[generalized gradient approximation (GGA)] exchange-correlation 
approximations with respect to Co on graphene. Wehling \emph{et al.}\cite{Wehling2010,Wehling2011} studied this system by using the 
GGA+$U$ approach and showed that the
results depend qualitatively on the on-site Coulomb repulsion parameter $U$. In particular, the application of $U=4$ eV leads to
a rearrangement of the ground-state electronic configuration of Co in such a way that its spin state changes
from $S=1/2$ to $3/2$ due to the $3d\rightarrow4s$ promotion. Besides, as the $U$ 
parameter increases, the adsorption energy of the adatom becomes significantly lower, leading to a larger Co--graphene separation. 
Similar results were reported by Chan \emph{et al.}\cite{Chan2011a}
The employment of hybrid functionals that incorporate some fraction of the exact nonlocal exchange also leads to qualitatively similar
results, as demonstrated by Jacob \emph{et al.}\cite{Jacob} It was also shown in Ref.~\onlinecite{Jacob} that the inclusion of the Co--graphene 
hybridization in terms of a many-body theory (one-crossing approximation method) significantly affects the splittings of the $3d$ cobalt orbitals in the crystal field of
graphene.
All these results indicate a key role of electronic correlations in the determination
of the ground-state properties of the Co--graphene system, and they suggest that more systematic and reliable approaches should be used in order
to properly describe electronic properties and chemical bonding of Co on graphene. In turn, accurate calculations of these 
properties are necessary to judge how stable Co impurities on graphene are, and how the properties of graphene itself are affected by the
impurities.

In contrast to previous first-principles studies, in the present paper a wave-function-based approach to investigate a Co adatom on graphene is employed. 
We use the multiconfigurational complete active space self-consistent field formalism (CASSCF) along with second-order perturbation theory,
which allows us to treat static (strong) and dynamical (weak) electron correlations in a balanced way. The main advantages this method offers compared to the DFT are: 
(i) systematic treatment of the exchange-correlation effects, (ii) proper electronic configuration of free TM atoms, (iii) the presence 
of dispersive interactions, (iv) correct asymptotic behavior in the dissociation limit for open-shell systems, and (v) the possibility to study excited states.
The price to be paid is the fact that the applicability of this method is currently limited only to finite and relatively small systems. As a result, it is
necessary to model the infinite graphene lattice by a finite cluster, approximated in this work by a coronene molecule (C$_{24}$H$_{12}$). This approximation 
looks reasonable if the \emph{local} adatom-substrate interactions dominate in the system. Similar cluster approaches have been used so far to 
investigate local correlation effects in extended systems by means of the multireference quantum chemical methods (see, e.g., 
Refs.~\onlinecite{Sadoc,Sharifzadeh,Hozoi}).

We show that there are two binding mechanisms of Co on graphene, which lead to different electron configurations of Co (high-spin $3d^74s^2$ and low-spin $3d^9$).
The first arises from the weak dispersive interaction (physisorption), while the second comes from the strong orbital hybridization (chemisorption). Despite the chemisorption, 
the adsorption energies are small and do not exceed 0.35 eV in both cases, which is substantially lower than the binding predicted in previously reported studies. 
We analyze the results using a simple model Hamiltonian, and we conclude that the systematic account of the electron correlations is crucial for 
a proper description of the adsorption of TM atoms on graphene. Based on this fact, we discuss the limitations and applicability of the DFT-based approaches to 
the TM/graphene systems.

The paper is organized as follows.
In Sec.~II we present the structural model, we briefly describe the CASSCF method, and we present other details on the calculations. 
Section III is devoted to the results obtained from first-principles calculations, 
where we first consider a single Co atom as a benchmark (Sec.~III~A), and then it describes the adsorption of Co on graphene (Sec.~III~B). 
The nonequivalence of the 3$d$ orbitals of Co and their influence on the adsorption are discussed in Sec.~III~C. 
In Sec.~IV, a mean-field analysis of the CASSCF results is performed. In Secs.~IV~A and IV~B we present a model Hamiltonian and discuss how the parameters of this Hamiltonian
can be determined, respectively. In Sec.~IV~C, a discussion on the calculated electron-electron interaction parameters is given. The correspondence with previous 
theoretical studies and the limitations of the DFT-based approaches with respect to the TM atom adsorption on graphene are discussed in Sec.~IV~D.
In the last section (Sec.~V), we briefly summarize our results and conclude this study.

\section{Calculation details}
\subsection{Structural model}

Many unique physical properties of graphene are determined by its gapless cone-shaped band structure, 
which arises as a result of the regular arrangement of carbon atoms in the honeycomb lattice. 
If translational symmetry of the lattice is broken, e.g., by contact with defects, impurities, or underlying surfaces, the electronic 
structure also changes, affecting the properties of graphene. Moreover, graphene should be spatially quite extended to retain its
properties. In particular, properties of polycyclic aromatic hydrocarbons (PAHs) are known 
to converge extremely slowly with the increase of the number of six-membered carbon rings. Theoretical estimations show that the 
highest occupied molecular orbital-lowest unoccupied molecular orbital (HOMO-LUMO) gap in large PAH molecules closes only at a cluster size of $\gtrsim$ 1000 benzene rings.\cite{Forte}

However, by investigating the local phenomena, such as, for example, interactions with atoms (molecules), an exact description of graphene's 
electronic structure is not necessary. Bonding caused by covalent interactions is essentially local and determined primarily by the atomic arrangement as well as by the
interatomic interactions in the host structure. These interactions vanish very rapidly with distance and thus depend only on
the local surroundings of an impurity. Therefore, it appears reasonable to consider only a local part of graphene for modeling this type
of interaction. Another aspect of strong chemical bonding is the ionic interactions, which scale linearly with distance 
for an unscreened case. In this situation, not only do neighboring atoms influence the bonding, but a rather large cluster is 
required to describe the interaction in the impurity-substrate system. Ionic interactions occur mainly in the case of charge transfer 
between an impurity and a surface. As we will show below, although in the case of maximal
3$d$ shell filling (3$d^9$) some electron transfer from Co to graphene is expected, the charge transfer is
quite small and any significant role of the distant surface atoms can be ruled out. A similar reasoning can be applied to the van der Waals (vdW) \mbox{-like} interaction
due to its weak strength.

\begin{figure}[!tbp]
\includegraphics[width=0.40\textwidth, angle=0]{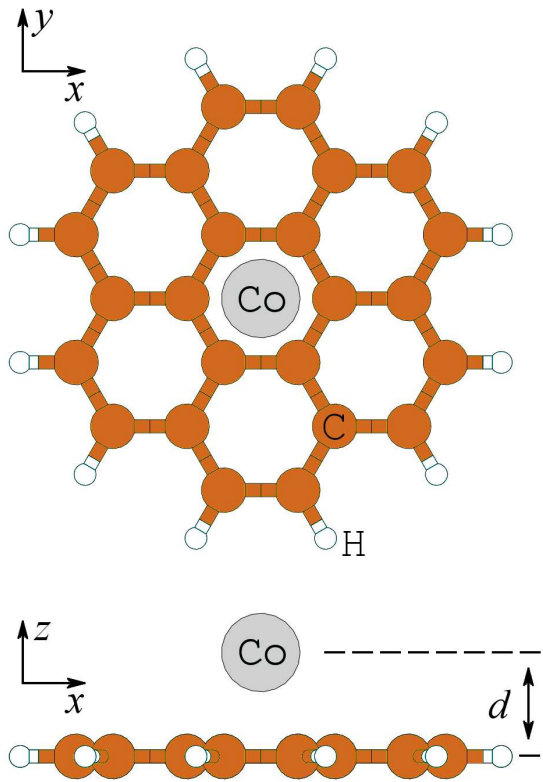}
\caption{(Color online) Schematic representation of cobalt supported on a coronene molecule used in this work as an approximant to the local structure of graphene.}
\label{coronene}
\end{figure}

The aforementioned arguments allow us to model the surface of graphene by using the finite cluster approach. More specifically, we 
consider a molecule of coronene (C$_{24}$H$_{12}$) to be a cluster for the local graphene structure (Fig.~\ref{coronene}). Similar finite models
have been used so far in studies on the physisorption of atoms (molecules) on graphene (see, e.g., Refs.~\onlinecite{Sheng1,Voloshina}). 
To make the cluster more realistic, the C--C bond lengths of coronene were adjusted to the bond length of graphene (1.42 \AA).
Due to symmetry reasons as well as for the sake of computational simplicity,
we consider only the center of this cluster as an adsorption site for the Co adatom. This site corresponds to a hollow site ($\eta^6$ coordination)
on graphene and has been repeatedly predicted to be the most favorable adsorption site for Co on undistorted graphene,
according to previous DFT studies. The same adsorption site is predicted for the cobalt-coronene complex.\cite{Kandalam} 
It is worth noting that relaxation of graphene may lead to the change of the most stable position of 
cobalt toward a top site, as has been shown in the GGA+$U$ studies.\cite{Wehling2010,Chan2011a} However, the energy difference between hollow and top 
sites was predicted to be quite small. In our study we do not take the effects of surface relaxation into account and we assume that graphene is locally flat. 

Due to the $C_{6v}$ point symmetry, the $3d$ orbitals of cobalt supported on a hollow site of graphene split into a single 
degenerate $A_1$ orbital, and two doubly degenerate $E_1$ and $E_2$ orbitals, derived from the $d_{z^2}$, $d_{xz}$ ($d_{yz}$), and
$d_{xy}$ ($d_{x^2-y^2}$) atomic orbitals, respectively. Since the $\pi$ orbitals of graphene transform 
according to the $E_1$ and $E_2$ representations in the vicinity of the Fermi level, the hybridization between the cobalt orbitals of 
$A_1$ symmetry ($3d_{z^2}$, $4s$) and graphene $\pi$ orbitals is suppressed.

\subsection{Method}

The starting point of an \emph{ab initio} quantum chemical treatment is the HF approximation. The HF method is able to describe only the electrostatic and exchange 
contributions to the electron-electron interaction energy, whereas correlation effects are not taken into account, $E_{\mathrm{HF}}=E_{\mathrm{total}}-E_{\mathrm{corr}}$. 
A hierarchy of widely used post-HF approaches [many-body perturbation theories, truncated 
configuration interaction (CI), and coupled cluster (CC) methods]\cite{Helgaker} is applicable primarily to the closed-shell systems, where correlation effects are 
rather weak and do not significantly affect the electronic structure of the system under investigation. 
This type of correlations is usually referred to as \emph{dynamical} correlations,\cite{Note1} according to the nomenclature commonly used in the quantum chemical literature. 
On the other hand, strong correlations may substantially affect electronic structure, so that the initial HF wave function (single Slater determinant) becomes unreliable 
and not suitable for further corrections. This is often the case for open-shell systems with near-degeneracies, including TM compounds. 
Effects that cannot be properly described at the single-determinant level are usually referred to as \emph{static} correlations. Total correlation energy can thus be given as the 
sum of the static and dynamical terms, $E_{\mathrm{corr}}=E_{\mathrm{stat}}+E_{\mathrm{dyn}}$.\cite{Mok,Hollett}

One of the efficient \emph{ab initio} approaches that allows systematic treatment of the static correlations is the CASSCF method.\cite{Roos_CAS}
In this method, the wave function is constructed as a full CI expansion within a limited set of
\emph{active} orbitals, for which all possible occupations are allowed. The other orbitals (inactive) have occupation numbers 
exactly equal to 2 and are treated at the HF level. Importantly, not only expansion coefficients but also the
orbitals are simultaneously optimized in the CASSCF, which makes this method quite flexible.
Dynamical correlations can be taken into account on top of the CASSCF wave function using multireference perturbation theories.
In this work, we employ the nonrelativistic CASSCF method in conjunction with the strongly contracted variant of the second-order $N$-electron 
valence state perturbation theory (NEVPT2)\cite{NEVPT2} as implemented in the {\sc orca} program package.\cite{orca}

In all calculations we employed the Ahlrichs triple-$\zeta$ valence basis set (TZV)\cite{Ahlrichs1} for all atoms. The cobalt atom and its six
neighboring carbon atoms (central ring in the coronene molecule) were additionally supplemented by (2$df$)- and ($d$)-polarization functions (TZVPP and TZVP basis sets), respectively.\cite{Ahlrichs2} 
We did not apply any basis set superposition error (BSSE) corrections to the calculated energy values.
The counterpoise correction (CP),\cite{BB} widely used in highly accurate quantum chemical calculations, does not usually lead to 
significant changes in the interaction energies for weakly bounded systems, whereas its application to strongly bounded systems is
controversial.\cite{Wright,Sheng2} Moreover, the reliability of the CP with respect to the multiconfiguration methods appears questionable.
As a starting point for the CASSCF calculations, we used the natural orbitals obtained at the level of unrestricted single-reference 
M\o ller-Plesset second-order perturbation theory (MP2). The calculated energies and orbital gradients were converged within 10$^{-8}$ and 10$^{-4}$ a.u., respectively.
It should also be noted that upon evaluation of four-index integrals, we \emph{did not} use any 
simplifications such as, e.g., the resolution of the identity approximation (RI).\cite{Neese2003}

\subsection{Active space}

\begin{figure}[!tbp]
\includegraphics[width=0.48\textwidth, angle=0]{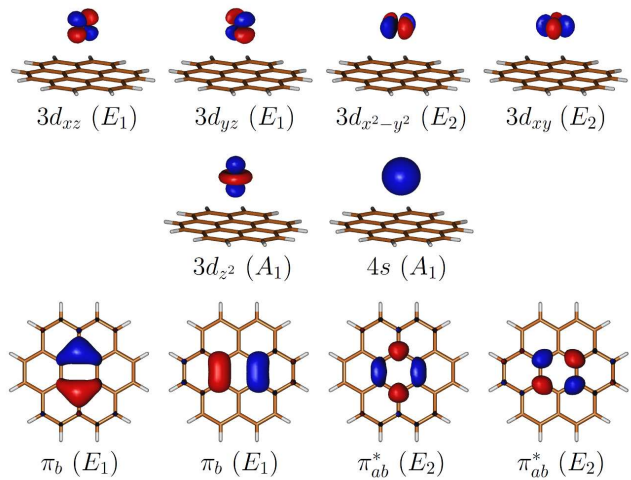}
\caption{(Color online) The set of orbitals included in the active space within the CASSCF approach (in the form of natural orbitals). The symmetry of the orbitals is shown in parentheses.}
\label{act_space}
\end{figure}

The active space used in this work for the CASSCF calculations consists of five 3$d$ Co orbitals, one 4$s$ Co orbital, and two pairs of bonding ($\pi_{b}$) and antibonding ($\pi_{ab}^*$) orbitals 
of the coronene molecule. This gives rise to 13 correlated electrons in 10 active orbitals [CAS(13,10)]. In Fig.~\ref{act_space}, we show schematically the active space in the form of natural orbitals 
obtained at the CAS(13,10) level for the ground state of Co at the distance of $d=3.1$ \AA \ from the surface.

By selecting the surface orbitals, we were guided by the following criteria: (i) the orbitals should be localized in the center of the surface, i.e., in spatial proximity to the Co atom; 
(ii) the orbitals should be either of $E_1$ or $E_2$ symmetry, since the contribution of the orbitals of $A_1$ symmetry is not likely in the graphene-impurity bonding; (iii) the orbitals should
be close to the Fermi level and represent a bonding-antibonding pair in order to mimic the $\pi$ and $\pi^*$ orbitals of graphene. The chosen surface orbitals obey these criteria. 
Moreover, except for the $A_1$ metal orbital, for all the other metal orbitals there is a corresponding ligand pair of the same symmetry. It is worth mentioning that the shape of
the selected $\pi$ orbitals of coronene is similar to the maximally localized Wannier functions associated with the $\pi$ orbitals of graphene.\cite{Mokrousov}

We note that in order to avoid unreasonably expensive calculations, our choice of active orbitals does not include explicitly the 
3$d$ double shell effect, which is still needed for an accurate treatment of the excitation energies in TM atoms.\cite{Andersson}
As we will show, however, the size of the chosen active space is sufficient to obtain a reasonable agreement between calculated and
experimental values of the $3d\leftrightarrow4s$ promotion energies for a single Co atom.

\section{First-principles results}

\subsection{Single cobalt atom}

The filling of the $d$-shells of TM atoms is known to occur in competition between the $d$ and $s$ states. As a result, low-lying excitations in the spectrum of a single atom are composed exclusively 
of the 3$d$~$\leftrightarrow$~4$s$ 
transitions. This competition is more pronounced upon formation of TM complexes where TM--ligand bonding takes place. While all first-row TM atoms accommodate electrons in the 4$s$ shell, the 
occupation of this shell becomes much less favorable upon contact with a ligand due to the Pauli repulsion
between the ligand and the very diffuse 4$s$ orbital of the TM atom. Therefore, for investigating the adsorption of TM atoms, it is of great importance to take into account the interplay between different
electronic states. Besides the general consideration, the important role of the 3$d$ and 4$s$ orbitals in the interaction of Co with graphene has been shown by previous 
studies.\cite{Wehling2010,Jacob,Wehling2011}

To investigate the role of different electronic states in adsorption of Co on graphene, we consider five different electronic configurations of Co:
(a) 3$d^7$4$s^2$ ($S$=3/2), (b) 3$d^8$4$s^1$ ($S$=3/2), (c) 3$d^7$4$s^2$ ($S$=1/2), (d) 3$d^8$4$s^1$ ($S$=1/2), and (e) 3$d^9$4$s^0$ ($S$=1/2). As one can see, 
these states correspond to two quartet and three doublet states with different occupancies of the 3$d$ and 4$s$ Co shells. 
The experimental ground state of a single cobalt atom corresponds to a high-spin quartet state, 3$d^7$4$s^2$. Unlike the LDA/GGA results obtained within the DFT,\cite{Moroni} 
this state can be predicted even at the Hartree-Fock level of theory. However, in order to obtain reasonable excitation energies, a multireference
treatment is required.\cite{Botch} 

In Table \ref{table1}, we show the excitation energies of a single Co atom calculated at the NEVPT2 level for the considered electronic states.
These states can be assigned to the lowest atomic terms of the corresponding configurations in the experimental spectrum as follows (in ascending order):  
$a^4F$ (ground), $b^4F$, $a^2F$, $a^2G$, and $c^2D$.\cite{Pickering} 
The energies of these terms, averaged over the spin-orbit components $m_J$, are given in Table \ref{table1}. The states $b^4F$ and $a^2F$ correspond to the two low-lying excited states of Co, 
while the others ($a^2G$ and $c^2D$) relate to much higher excitations. One can see that the calculated energies of the first two excited states are in very good agreement with experiment, 
while for the third excited state the agreement is slightly worse. The maximum discrepancy can be seen in the case of 
the highest energy level. This is not surprising in view of the fact that the active space used in this work is restricted only to the 3$d$ and 4$s$ orbitals,
whereas higher orbitals are necessary to properly describe the high-energy excitations. Another necessary condition for an accurate prediction of the excited energy levels
is relativistic corrections,\cite{Autschbach} which were also not utilized in the present work.
Nevertheless, despite some quantitative discrepancies, the agreement obtained between theoretical and experimental spectra for a single Co 
atom can be considered
as satisfactory. The consistency achieved with experimental data shows the reliability of the computational approach as well as an adequate choice of parameters for studying the Co--graphene system.

    \begin{table}[!tbp]
    \centering
    \caption[Bset]{Calculated and experimental excitation energies of a single Co atom corresponding to transitions from the ground (3$d^7$4$s^2$, $S$=3/2) to low-lying excited states associated 
with the electronic configurations considered in this work. Experimental energies correspond to the $m_J$ averaged value of the lowest atomic term of a given configuration (Ref.~\onlinecite{Pickering}).}
    \label{table1}
\begin{ruledtabular}
 \begin{tabular}{ccccccc}
%\hline
                      &  \multicolumn{2}{c}{$S$=3/2} & \multicolumn{3}{c}{$S$=1/2} \\
\cline{2-3}
\cline{4-6}
                      &  3$d^7$4$s^2$   &  3$d^8$4$s^1$  & 3$d^7$4$s^2$  &  3$d^8$4$s^1$  &  3$d^9$4$s^0$ \\
     \hline
$\Delta E_{calc}$, eV &       0.0       &       0.38     &     2.21      &      0.82      &      4.30     \\
$\Delta E_{exp}$, eV  &       0.0       &       0.41     &     1.96      &      0.86      &      3.35     \\
  Atomic term         &     $a^4F$      &     $b^4F$     &     $a^2G$    &      $a^2F$    &     $c^2D$    \\
    \end{tabular}
\end{ruledtabular}
    \end{table}

\subsection{Adsorption of cobalt}

To reproduce the potential energy curves, we perform a series of state-specific calculations for 
different separations between Co and graphene. We define the adsorption energy of Co in the state $i$ as
\begin{equation}
E_{\mathrm{ads}}^{i}(d)=E_{\mathrm{surf+Co}}^{i}(d)-E_{\mathrm{surf}}-E_{\mathrm{Co}}^{\mathrm{ground}},
\label{eq_ads}
\end{equation}
where $E_{\mathrm{surf+Co}}^{i}(d)$ is the energy of the interacting system in the $i$ state of Co separated by the distance $d$ from the surface, $E_{\mathrm{surf}}$ is the energy of the 
clean surface, and $E_{\mathrm{Co}}^{\mathrm{ground}}$ is the energy of the isolated Co atom in its ground state. Equilibrium adsorption energies for different Co states can be obtained by 
taking the minimum of the corresponding adsorption energy functions, $E^i_{\mathrm{ads}}(d_{\mathrm{eq}})=\mathrm{min}[E_{\mathrm{ads}}^i(d)]$. Since the NEVPT2 approach employed in this work is
size-consistent, the dissociation limit can be correctly reproduced, and at large distances Eq.~(\ref{eq_ads}) gives the interconfigurational energies between the ground and excited states of a single 
Co atom, $E_{\mathrm{ads}}^{i}(d \rightarrow \infty)=E^i_{\mathrm{Co}}-E^{\mathrm{ground}}_{\mathrm{Co}}$.

\begin{figure}[!btp]
\vspace{0.2cm}
\includegraphics[width=0.48\textwidth, angle=0]{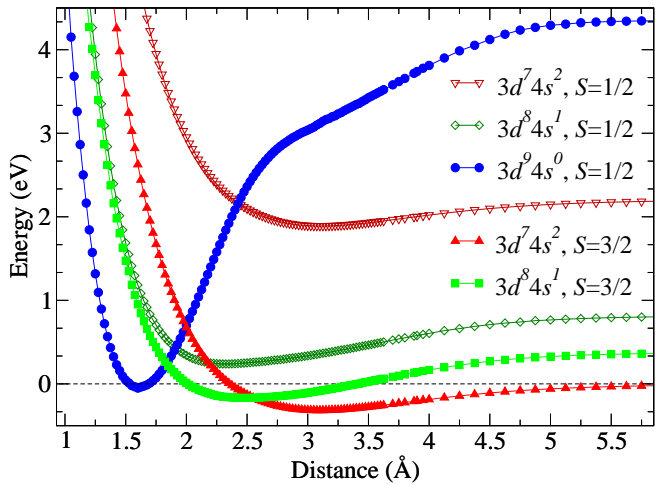}
\caption{(Color online) Potential energy curves for different electronic configurations of Co on graphene. Zero energy corresponds to the ground-state energy of the noninteracting system.}
\label{energies}
\end{figure}

In Fig.~\ref{energies}, we show the lowest adsorption energy curves calculated at the NEVPT2 level for different configurations of the Co adatom supported on graphene. The adsorption energies,
equilibrium distances, and orbital occupations for each configuration are summarized in Table \ref{table2}.
As expected, the shapes of the curves as well as equilibrium distances do not depend significantly on the spin and depend mainly on the 
orbital occupancies.  In the case of the same occupations, the low-spin curves lie higher in energy due to the difference in 
exchange energy. In the short-distance range, this difference gradually decreases, which can be associated with the loss of the $d-d$ exchange
energy effect that is known for molecular systems containing TM atoms.\cite{Yarkony} 

    \begin{table*}[!tbp]
    \centering
    \caption[Bset]{Calculated adsorption energies, equilibrium distances, and orbital occupations for Co on graphene in different electronic configurations. Occupation numbers correspond to the
equilibrium distances and are given in terms of L\"{o}wdin's charges\cite{Lowdin1950} and natural orbital occupations.\cite{Lowdin1955}}
    \label{table2}
\begin{ruledtabular}
 \begin{tabular}{ccccccc}
                         &  \multicolumn{2}{c}{$S$=3/2} & \multicolumn{3}{c}{$S$=1/2} \\
\cline{2-3}
\cline{4-6}
                         &  3$d^7$4$s^2$   &  3$d^8$4$s^1$  & 3$d^7$4$s^2$  &  3$d^8$4$s^1$  &  3$d^9$4$s^0$ \\
     \hline
$E_{\mathrm{ads}}$, eV   &      -0.32      &      -0.17     &     1.89      &       0.24     &      -0.05       \\
$d_{\mathrm{eq}}$, \AA   &       3.1       &       2.5      &      3.1      &        2.3     &       1.6        \\
Orbital occupations\footnotemark[1]   &   3$d^{7.00}$4s$^{1.89}$   &  3$d^{7.98}$4$s^{0.87}$  &    3$d^{7.00}$4s$^{1.89}$     &   3$d^{7.98}$4$s^{0.86}$   &    3$d^{8.22}$4$s^{0.06}$      \\
Orbital occupations\footnotemark[2]   &   3$d^{7.0}$4s$^{2.0}$   &  3$d^{8.0}$4$s^{1.0}$  &    3$d^{7.0}$4s$^{2.0}$     &   3$d^{7.0}(sd)_{+}^{0.9}(sd)_{-}^{1.1}$\footnotemark[3] &    3$d^{8.7}$4$s^{0.0}$      \\
    \end{tabular}
\end{ruledtabular}
\footnotetext[1]{Based on L\"{o}wdin's population analysis.\cite{Lowdin1950}}
\footnotetext[2]{Based on natural orbital analysis.\cite{Lowdin1955}}
\footnotetext[3]{In this case, the 3$d_{A_1}$ Co orbital slightly hybridizes with the 4$s$ orbital forming a $3d$--$4s$ hybrid, $(sd)_{\pm}=1/\sqrt{2}(3d_{A_1}\pm4s)$.}
    \end{table*}

The adsorption of transition metals differs from the adsorption of monovalent or closed-shell adsorbates.
As can be seen from Fig.~\ref{energies}, the ground-state electronic structure of Co undergoes several transitions upon adsorption. In turn, the adsorption energy curve (the minimum energy curve) is 
composed of three different potential energy curves. As the atom approaches the surface, its most favorable configuration (quartet, $3d^74s^2$)
changes to $3d^84s^1$ (quartet) at a distance of $\sim$2.6 \AA, and to $3d^94s^0$ (doublet) at shorter distances ($\sim$1.8 \AA). This behavior can be qualitatively understood by noting 
that in the vicinity of the ligand, the occupation of the $4s$ orbital becomes less favorable due to the Pauli repulsion, while the occupation of the $3d$ orbital becomes more favorable due to the screening
of the Coulomb repulsion in the 3$d$ shell of Co.
We analyze these effects in more detail in Sec.~IV~C.
The energies of the two other doublet configurations lie higher over the entire range of distances. The absolute values of the adsorption energies are quite small, reaching 0.32 eV in the 
high-spin $3d^74s^2$ case, while in the other cases the values are even smaller (see Table \ref{table2}). These values are significantly smaller than the adsorption energy predicted by the DFT
studies ($\sim$1--2 eV),\cite{Wehling2010,Mao,Johll,Valencia,Cao,Liu,Ding,Wehling2011,Lima} but are consistent with experimental observations of the fast surface diffusion of metal atoms on 
graphene,\cite{Manzo} which suggest weak bonding in the system.

Although the minimum energy curve shown in Fig.~\ref{energies} exhibits three extrema, only two of them (low-spin, $3d^94s^0$ and high-spin, $3d^74s^2$) are well separated and can be 
regarded as stable configurations. The third minimum (high-spin, $3d^84s^1$) corresponds to the metastable state.
Among the two stable configurations, the high-spin minimum lies lower in energy than the low-spin one and, therefore, represents the global minimum. As the low-spin configuration is less 
favorable in absolute values, there is an energy barrier to reach this state. On the other hand, if the adatom is trapped in the low-spin state, another barrier 
has to be overcome to return in the high-spin configuration. The energy barriers can be estimated 
as a difference between the transition state and high-spin (low-spin) state energies. As can be seen from Fig.~\ref{energies}, there is a transition state at a distance of 1.85~\AA~with an energy of
0.24 eV. Taking this state into account, we obtain the energy barriers of 0.56 and 0.29 eV, respectively for the high-spin$\rightarrow$low-spin and low-spin$\rightarrow$high-spin transitions.
We note that a qualitatively similar multiple-minima structure of the potential energy curves has been reported for graphene adsorbed on metal surfaces.\cite{Olsen,Mittendorfer,Kozlov}

Let us analyze the nature of the Co--graphene interactions.
The maximum adsorption energy obtained for Co (0.32 eV) is comparable with adsorption energies typical for small closed-shell molecules adsorbed on graphene,\cite{Rudenko2010} where only 
weak vdW interaction plays a dominant role. In fact, the same mechanism is responsible for the adsorption of Co in the $3d^74s^2$ and $3d^84s^1$ configurations. 
In these cases, the hybridization between the localized 3$d$ orbital of Co and the $\pi$ orbital of graphene is suppressed due to the Pauli repulsion between graphene and the diffuse 4$s$ Co orbital.
Interestingly, however, the surface repulsion favors some hybridization between the 3$d_{A_1}$ and 4$s$ Co orbitals in the low-spin $3d^84s^1$ state, as indicated in Table \ref{table2}.
In the cases where the 4$s$ orbital is occupied, the absence of the hybridization with the surface can be concluded by analyzing the occupancies of orbitals associated with the $3d$ Co orbital.
L\"{o}wdin population analysis\cite{Lowdin1950} as well as the occupation numbers of the natural orbitals\cite{Lowdin1955} show that at equilibrium distances, the Co orbitals do not have contributions 
from/to the $\pi$ orbitals of graphene, so that the corresponding occupation numbers are close to integer values (see Table \ref{table2}). 
It is also worth noting that the binding of Co with the nonempty 
$4s$ orbital occurs only at the level of second-order perturbation theory and not in the pure CASSCF calculations, which also confirms that the interaction between Co with a nonempty 4$s$ 
shell and graphene is of the weak vdW-type.

\begin{figure}[!btp]
\includegraphics[width=0.48\textwidth, angle=0]{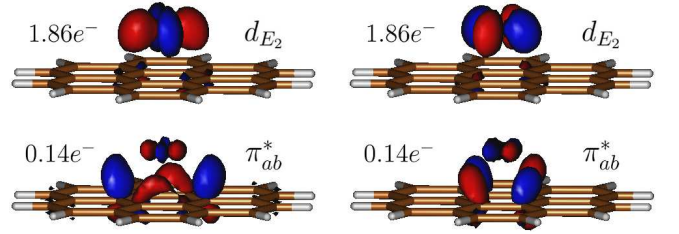}
\caption{(Color online) Isosurfaces of the natural orbitals associated with the hybridized $d_{E_{2}}$ (adatom) and $\pi_{ab}^{*}$ (surface) orbitals. The numbers correspond to the natural orbital 
occupation numbers.}
\label{nat_orbs}
\end{figure}

The situation with the low-spin configuration ($3d^94s^0$) is different. In this case, as can be seen in Fig.~\ref{energies}, the potential energy curve exhibits a marked linear dependence 
within the distance range of 2.8$-$4.7~\AA. This behavior suggests an ioniclike interaction between graphene and Co in the $3d^94s^0$ state. L\"{o}wdin population analysis shows that in this range,
the occupation of orbitals changes from $3d^{8.85}$ to $3d^{8.99}$, which means that some fraction of Co electrons is transferred toward graphene. This
can be understood in terms of the ionization energy lowering associated with the filling of the 3$d$ orbital. Therefore, the observed ioniclike interaction can be explained as a result of a minor charge 
doping.
At shorter distances, the $3d^94s^0$ adsorption curve exhibits a deep minimum typical for strong covalent interactions. In this situation, the hybridization between the 3$d$ Co and $\pi$ surface orbitals 
is more likely since the 4$s$ Co orbital is empty and the atom--surface repulsion is much weaker. 
As can be seen from Table \ref{table2}, the natural orbitals of Co in the $3d^94s^0$ state have noninteger occupation numbers, which suggests that there is an additional, 
many-body contribution to the Co--graphene bonding. 
Indeed, analysis of the wave function shows that the weight of the principal contribution in the CASSCF expansion, $\left| \pi_b^{4} d_{A_1}^{2}d_{E_1}^3d_{E_2}^4 \pi_{ab}^{*0}\right>$, is only 0.84. 
The remaining part of the wave function is distributed over many configurations, which mainly corresponds to the hopping of electrons from the $3d_{E_2}$ Co orbital to the $\pi_{ab}^*$ surface orbital.
Only one of these configurations reaches a contribution larger than 1\%, namely,  $\left| \pi_b^{4} d_{A_1}^{2}d_{E_1}^3d_{E_2}^1d_{E_2}^1  \pi_{ab}^{*1} \pi_{ab}^{*1} \right>$ (4\%). 
This allows us to conclude that the deep
minimum observed in Fig.~\ref{energies} for the adsorption of Co in the $3d^94s^0$ configuration is attributed to the donation of the metal electrons to the ligand.
The donation effect can be more clearly seen in Fig.~\ref{nat_orbs}, where we show isosurfaces of the natural orbitals associated with the Co 3$d_{E_2}$ and $\pi_{ab}^*$ surface orbitals.
Since the donation process involves orbitals of specific symmetry ($3d_{E_2}$), the degeneracy of the 3$d$ Co orbitals is broken and an energy splitting is expected between
different 3$d$ orbitals. We discuss this issue in the next section. 
It is also important to emphasize that the effect of considerable energy lowering of the 3$d^9$4$s^0$ Co state has an essential multiconfigurational character
and thus cannot be properly described at the single-determinant level.

\subsection{3$d$-orbital splitting}

Up to now, we have not paid any attention to the splitting of the 3$d$-orbitals, assuming that the electrons occupy the lowest energy states in a given configuration. 
It is clear, however, that when a TM atom approaches the surface, the crystal field tends to split the 3$d$-orbitals more strongly the smaller the distance. 
Moreover, symmetry-specific hybridization discussed in the previous section suggests that these
effects are not negligible. As we already mentioned, if a TM atom is 
deposited on a hollow site of graphene, its otherwise degenerate 3$d$-orbitals split into three groups according to the $A_1$, $E_1$, and $E_2$ irreducible representations of the $C_{6v}$ point symmetry.
In what follows, we assume that the 3$d$-orbital splitting does not depend on the particular Co state, i.e., the same for different electronic configurations.

\begin{figure}[!tbp]
\vspace{0.2cm}
\includegraphics[width=0.48\textwidth, angle=0]{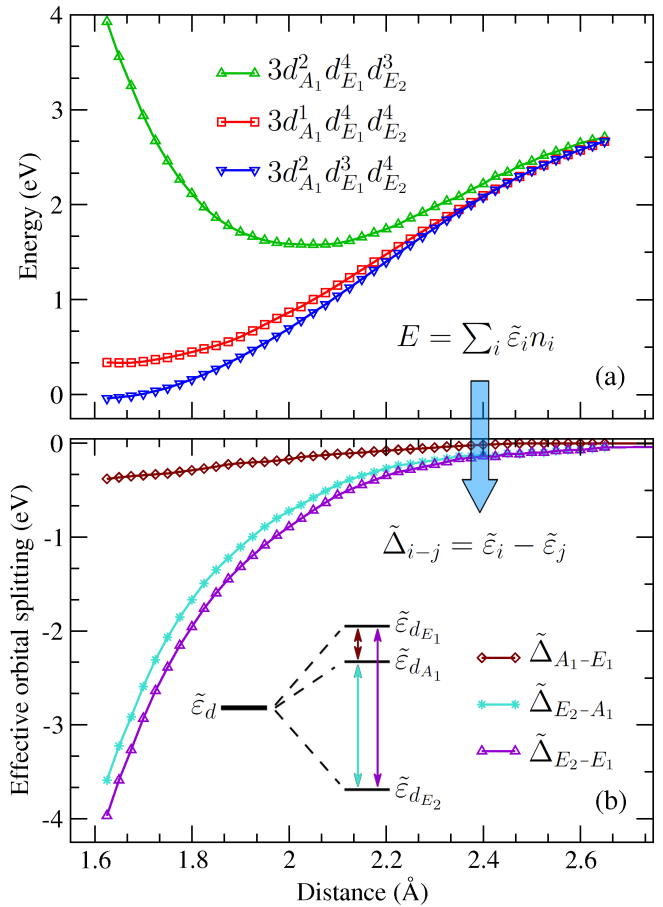}
\caption{(Color online) (a) Potential energy curves for different orbital occupations of Co
in the 3$d^9$ configuration; (b) energy differences between nonequivalent 3$d$-orbital levels of Co (the effective crystal-field splitting parameters) calculated as functions of the adatom--surface 
separation (see text for details). The splitting of the energy levels is shown schematically inside the lower panel.}
\label{splitting}
\end{figure}

To estimate the 3$d$-orbital splitting, we perform additional energy calculations for the 3$d^9$4$s^0$ configuration of Co, which correspond to different occupations of symmetry nonequivalent 
orbitals ($d_{A_1}$, $d_{E_1}$, and $d_{E_2}$). The potential energy curves for these states are given in Fig.~\ref{splitting}(a). One can see that at distances larger than 2.5 \AA, the energies are
almost identical, i.e., they do not depend on the particular occupation of the 3$d$ orbital. As the separation between Co and the surface decreases, different orbital occupations result in different
energies. This reflects the fact that the Co orbitals interact differently with the surface, which is also clear from the hybridization between the 3$d_{E_2}$ Co and
$\pi^{*}_{ab}$ surface orbitals, as discussed above. As a result of hybridization, the states with fully occupied 3$d_{E_2}$ orbitals are much more favorable, as can be seen from Fig.~\ref{splitting}(a).

Energy differences between the states of different 3$d$ orbital occupations can be interpreted in terms of the simple crystal-field picture. If we express the energy of a particular state as a sum of
one-particle energies, 
\begin{equation}
E=\sum_{i}\tilde{\varepsilon}_in_i,
\end{equation} 
the effective crystal-field splitting parameters, $\tilde{\Delta}_{ij}=\tilde{\varepsilon}_i-\tilde{\varepsilon}_j$, can be easily estimated from the 
total energies of different orbital occupations. In Fig.~\ref{splitting}(b), we show these parameters as functions of distance. First, it should be noted that the orbital energy levels follow the order 
$\tilde{\varepsilon}_{E_2}<\tilde{\varepsilon}_{A_1}<\tilde{\varepsilon}_{E_1}$, which does not depend on the distance. This order is consistent with previous DFT\cite{Wehling2010,Valencia} and 
many-body\cite{Jacob} studies. As can be seen from Fig.~\ref{splitting}(b), the $\tilde{\Delta}_{A_1-E_1}$ splitting parameter is relatively small, not exceeding 0.5 eV even at short distances. 
On the other hand, the splitting between the $E_2$ and $E_1$ ($A_1$) orbitals is significantly larger, reaching 4 eV at a distance of 1.6 \AA. 

There are two main reasons that can account for the large splitting.
The first arises from the direct interaction of Co electrons with the surface, resulting in the hybridization and/or interorbital repulsion. The second can be attributed to the screening effects 
associated mainly with the reduction of the Coulomb repulsion in the Co 3$d$-shell. Indeed, as we have shown before, the 3$d_{E_2}$ Co orbital hybridizes with the surface, lowering the energy of this 
orbital. At the same time the hybridization causes delocalization of the electrons occupying this orbital (see Fig.~\ref{nat_orbs}), which in turn reduces the Coulomb repulsion and leads to further
energy lowering. In the next section, we analyze these effects quantitatively based on a simple model Hamiltonian.

\section{Mean-field analysis}

\subsection{Model Hamiltonian}

Analysis of a many-body wave function in terms of the usual single-particle quantities is challenging. Although the single-particle representation of many-body characteristics can in principle be constructed
(e.g., in the many-body Green's function formalism), the CASSCF wave function cannot be readily mapped onto the single-particle picture. As a result, a number of physically relevant quantities, such as,
for example, the density of states, are not available in the multireference methods. In this work, in order to more deeply understand the physics behind the adsorption of Co on graphene, we 
map the CASSCF results onto a mean-field model. To this end, we consider a simple Hamiltonian:
\begin{eqnarray}
\label{hamilt}
H=\sum_{i\sigma} \varepsilon_in_{i}^{\sigma} + \frac{1}{2}\sum_{i,\sigma} U_i n_i^{\sigma} n_i^{\bar{\sigma}} \quad \quad \quad \quad \quad \quad \nonumber \\
+ \frac{1}{2}\sum_{\substack{ij,\sigma \\ i \neq j}} \left(U'_{ij} - J_{ij}\delta_{\sigma \sigma'} \right) n_i^{\sigma} n_j^{\sigma'}. \quad
\end{eqnarray}
Here $i$ and $\sigma$ label orbital indices and spin-projections for 3$d$ and 4$s$ electrons of Co ($\bar{\sigma}$ implies spin
inversion), respectively, $\varepsilon_i$ is the energy of the $i$ electron in the absence of electron-electron 
interaction, and $n_i^{\sigma}$ is the particle number operator. The parameters $U_i$, $U'_{ij}$, and $J_{ij}$ determine
intraorbital and interorbital Coulomb interactions as well as interorbital exchange interaction (Hund's rule coupling), respectively. 
The first term in this 
Hamiltonian describes the one-electron contribution to the energy, including the interaction with the crystal field of the ligand. 
The rest represents the Coulomb interaction Hamiltonian in the Kanamori 
parametrization,\cite{Izyumov} which describes the electron-electron interactions. In this parametrization, the Coulomb and exchange
parameters for the 3$d$ electrons satisfy the relationship $U-U'=2J$ if the system is rotationally invariant.
It should be noted that although the given Hamiltonian does not explicitly contain atom--ligand interactions, these interactions are
implicitly taken into account in the first term through the crystal-field splittings of the Co orbitals. In turn, since the $U$, $U'$, 
and $J$ parameters are defined only for the Co part of the system, they should be considered as effective.
For the sake of simplicity, we also assume that the rotational invariance condition holds and the electron-electron interaction parameters have the same values for different 3$d$ orbitals 
(i.e., averaged over orbitals). 
Finally, we neglect the intraorbital Coulomb repulsion between electrons in the 4$s$ orbital 
($U_{4s}$), and interorbital 3$d$--4$s$ repulsions ($U'_{s-d}$), taking into account the delocalized character of the 4$s$ orbital. The interaction
between the 4$s$ and 3$d$ electrons is taken into account through the exchange interaction parameter $J_{s-d}$.

\subsection{Determination of model parameters}

Since the Hamiltonian given by Eq.~(\ref{hamilt}) is diagonal, the energies of different atomic states of Co can be readily
expressed in terms of the parameters $U_{d-d}$, $U'_{d-d}$, $J_{d-d}$, and $J_{d-s}$, and the 3$d$--4$s$ orbital energies
$\varepsilon_{A_1}$, $\varepsilon_{E_1}$, $\varepsilon_{E_2}$, and $\varepsilon_{4s}$. By comparing the energies of 
different states obtained in the model with the values obtained from the first-principles calculations, the magnitude of the 
electron-electron interaction parameters can be extracted if the crystal-field splitting parameters, $\Delta_{ij}=\varepsilon_i-\varepsilon_j$, are known (see the Appendix for the specific expressions).
Unfortunately, these parameters do not generally correspond to the effective $\tilde{\Delta}_{ij}$ parameters obtained in Sec.~III~C. As we already discussed, apart from the
direct interactions with the crystal field, the $\tilde{\Delta}_{ij}$ parameters may contain screening effects, enhancing the splitting due to additional lowering of the 3$d_{E_2}$ energy level.
However, there is no straightforward way to establish a precise quantitative relationship between these two effects from first-principles calculations.

The splitting in metal atoms due to the presence of neutral ligands is in general not large, especially for low-coordinated complexes such as atoms on surfaces. Taking this fact into account, 
it appears reasonable to employ significantly smaller splittings than $\tilde{\Delta}_{ij}$ upon determination of the parameters of the model Hamiltonian. To be more specific, we scale the effective splitting by a factor of 0.2, so that
$\Delta_{ij}=0.2\tilde{\Delta}_{ij}$. At the distance of $d=1.6$ \AA, this choice is close to the splitting obtained by means of the standard DFT-GGA calculations without using the $U$-corrections ($\Delta_{E_2-E_1}=0.8$ eV, $\Delta_{E_1-A_1}=0.5$ eV).\cite{Wehling2010}
The use of smaller scaling factors does not significantly affect the results, while the larger factors lead to an unphysical (nonmonotonic) dependence of $U_{d-d}$ with distance. This can be considered as yet 
another criterion for the physically reasonable choice of the scaling factor.
Finally, we note that the uncertainty in the determination of the crystal-field splitting affects only the Coulomb repulsion parameters in the short-distance range.

\begin{figure}[!tbp]
\vspace{0.2cm}
\includegraphics[width=0.49\textwidth, angle=0]{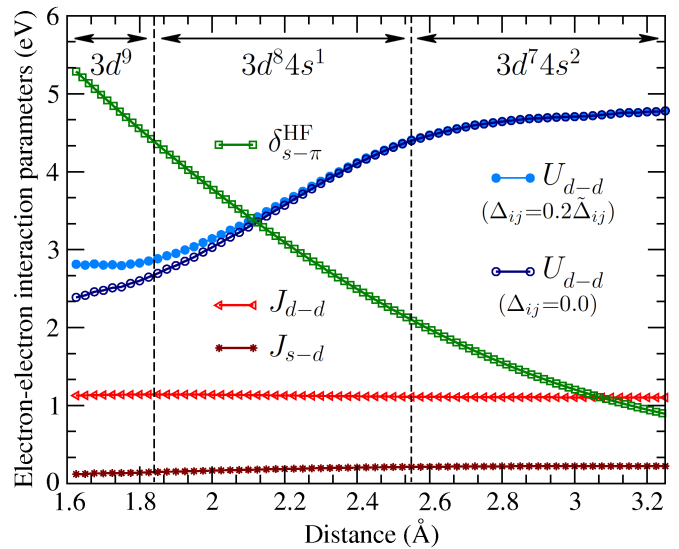}
\caption{(Color online) Electron-electron interaction parameters for Co on graphene as a function of distance: exchange couplings ($J_{s-d}$ and $J_{d-d}$) and intraorbital Coulomb 
repulsion ($U_{d-d}$). The latter is shown for different splitting parameters $\Delta_{ij}$ (see text for details). $\delta_{s-\pi}^{HF}$ quantifies the Pauli repulsion between
the 4$s$ (Co) and $\pi_b$ (surface) orbitals.\cite{delta-calc}
The dashed lines mark the most favorable 
electronic configuration of Co for a given range of distances (in accordance with potential energy curves given by Fig.~\ref{energies}).}
\label{parameters}
\end{figure}

\subsection{Analysis of the electron-electron interactions}

In Fig.~\ref{parameters}, we show the calculated electron-electron interaction parameters as functions of the Co-graphene 
distance.
As one can see, the exchange coupling parameters ($J_{s-d}$
and $J_{d-d}$) are almost constant and do not change over the entire range of distances. $J_{d-d}$ has a typical value commonly used in the LDA+$U$, and model Hamiltonian calculations of TM 
compounds,\cite{Grechnev} and it amounts to $\simeq$1.1 eV. 
As expected, the $s$-$d$ exchange interaction is significantly smaller (0.1--0.2 eV) and does not play an important role in determining the electronic structure. 
In contrast, the intraorbital Coulomb repulsion ($U_{d-d}$) significantly decreases for shorter distances (from 5.3 eV in the atomic limit to $2.6\pm0.2$ eV at a distance of 1.6 \AA).

As we have shown in Sec.~III~B the adsorption of Co atoms on graphene differs from the adsorption of closed-shell systems because it involves transitions between different electronic configurations.
The nature of these transitions can be interpreted in terms of the two primary effects: (i) Coulomb repulsion between the electrons in the 3$d$ shell of Co and (ii) electrostatic (Pauli) repulsion 
between the 
4$s$ (Co) and $\pi$ (graphene) electrons.  The first effect can be described by the $U_{d-d}$ parameter and its evolution with distance, while the second cannot be captured by the model 
Hamiltonian [Eq.~(\ref{hamilt})] since it does not explicitly contain the surface energy levels. Since the 4$s$--$\pi$ orbital repulsion is essentially an electrostatic effect,
it can be characterized by the 4$s$--$\pi$ orbital splitting obtained by means of the simple HF approach $\delta_{s-\pi}^{\mathrm{HF}}$,\cite{delta-calc} whose
variation with distance is also shown in Fig.~\ref{parameters}.

Analyzing the behavior of $U_{d-d}$ and $\delta_{s-\pi}$, one can see that both effects tend to move Co electrons from the 4$s$ to the 3$d$ shell, which leads to the interconfigurational transitions of 
the TM atom upon adsorption. At large distances, the variation of the two factors is small, and the most favorable configuration ($3d^74s^2$) is governed by the strong repulsion in the 3$d$ shell.
As the atom comes closer to the surface, the occupation of the 3$d$ shell becomes more favorable due to the reduction of the Coulomb repulsion, whereas the occupation of the 4$s$ shell becomes less
favorable due to the increasing Pauli repulsion between the 4$s$ and $\pi$ electrons. This gives rise to the $3d^84s^1$ and $3d^9$ configurations of Co.
 
It is important to emphasize that both described effects have a non-negligible contribution and play an important role in the determination of the ground-state electronic configuration of the adsorbate.
In comparison to the electrostatic repulsion between Co and graphene, the screening of the Coulomb repulsion is essentially a correlation effect and thus represents a challenge for
one-electron theories. In the next section, we analyze the perspectives of the DFT-based approaches in the description of TM atom adsorption on graphenelike structures.

\subsection{Connection to DFT}

As a final step, it is interesting to draw a parallel between the results obtained in the present work and the results of previously reported DFT studies.
The standard GGA calculations predict that the $3d^9$ configuration is the ground state of Co on graphene.\cite{Wehling2010,Wehling2011} 
The same configuration persists if a small $U$-correction ($U=2$~eV) is applied within the GGA+$U$ method.\cite{Wehling2010}
The application of larger corrections ($U=4$~eV) yields a configuration close to $3d^84s^1$.\cite{Wehling2010,Wehling2011} 
Finally, hybrid-functional calculations, which incorporate some fraction of the Hartree-Fock exchange energy, also predict
the $3d^84s^1$ configuration for Co on graphene.\cite{Jacob}

As one can see, the electronic configurations of
Co obtained within the DFT for small and intermediate values of $U$ (0--4 eV) are consistent with the configurations that correspond to the same values of $U$ in the present work 
(see Fig.~\ref{parameters} and the discussion above).
This agreement justifies the use of on-site Coulomb corrections in DFT for the description of TM atoms on graphene.
However, being significantly dependent on $U$, the DFT results are applicable only to a certain range of adsorption distances,
while the entire adsorption energy curve cannot be reproduced in a consistent way.\cite{Kulik} This should be carefully taken into account in further studies, especially in 
the determination of ground-state configurations.

The inability to describe the total potential energy curves for TM atoms on graphene within the DFT also leads to difficulties in the determination of the binding energies.
Since an energy calculation of an isolated TM atom requires its own $U$ correction to be reliable, the comparison of energies [e.g., by using Eq.~(\ref{eq_ads})] becomes meaningless.\cite{Kulik}
On the other hand, if the same $U$ is used for the calculation of energies of the interacting and noninteracting systems, the binding energies are strongly overestimated due to
an improper description of isolated TM atoms. Another point, though less important, that prevents DFT from giving accurate binding energies is the lack of
nonlocal dispersion interactions in the standard DFT approximations (LDA/GGA). 

In summary, the deficiency of the DFT-based approaches with respect to the description of TM atoms supported on graphenelike
structures can be attributed to their inability to handle electron correlations in a consistent and systematic way.

\section{Conclusion}

In this work, we have applied multireference quantum chemical methods to the systematic investigation of adsorption of a single cobalt atom on graphene.
It is found that the electron correlation effects play an important role in the adsorption. Far from the surface, these effects are not significant and manifest themselves only in the weak attractive 
interactions of the van der Waals type. In the vicinity of the surface, the role of correlations is more pronounced and can be attributed mainly to the screening of the strong Coulomb repulsion between 
electrons in the 3$d$ shell of Co. Due to the significant reduction of the Coulomb repulsion, the Co states with higher 3$d$ shell occupancy become more favorable at short distances. 
As a result, the total adsorption energy curve of Co on graphene is not monotonic and exhibits two minima, which correspond to the $3d^74s^2$ (high-spin) and $3d^9$ (low-spin) configurations.
Therefore, two stable states of cobalt on graphene are predicted even for a single (hollow) adsorption site.
 
We have shown that the calculated adsorption energies for Co/graphene are much smaller ($E_{\mathrm{ads}}<0.35$ eV) than was reported previously by DFT studies. The main reason is associated with
the inability to properly describe the total adsorption energy curve for TM adsorbates within the DFT. This, in turn, is related to the inconsistent treatment of the strong correlation effects, 
even in the presence of on-site Coulomb corrections (DFT+$U$ methods).

Although only the case of the Co adatom is considered in this work, we believe that the importance of the correlation effects is not specific only to the Co adsorption.
We expect qualitatively similar adsorption behavior for other TM atoms on graphene, particularly with respect to the interconfigurational transitions and relatively small binding energies. 
The determination of absolute values as well as the ground-state electronic configurations of different TM adatoms on graphene remains an open issue for further studies.

\section{Acknowledgments}

The authors thank Tim Wehling, Vladimir Mazurenko, and Danil Boukhvalov for helpful discussions. Support from 
the DFG Priority Program No.~1459 ``Graphene'' (Germany), from Stichting voor Fundamenteel Onderzoek der Materie (FOM, the Netherlands),
and from the Russian Scientific Program No.~12.740.11.0026 is gratefully acknowledged.

\appendix 

\section{Determination of electron-electron interaction parameters}

According to the Hamiltonian given by Eq.~(\ref{hamilt}), the energies of the five cobalt states of interest can be written as

\begin{eqnarray}
\label{a1}
E_{S=3/2}^{3d^74s^2} = 2\varepsilon_{A_1} + 2\varepsilon_{E_1} + 3\varepsilon_{E_2} + 2\varepsilon_{4s} \quad \quad \quad \nonumber \\
+ 2U_{d-d} + 11J_{d-d} + 7J_{s-d} + 19U'_{d-d},
\end{eqnarray}
 
\begin{eqnarray}
\label{a2}
E_{S=3/2}^{3d^84s^1} = \varepsilon_{A_1} + 4\varepsilon_{E_1} + 3\varepsilon_{E_2} + \varepsilon_{4s}  \quad \quad \quad \quad \nonumber \\
+ 3U_{d-d} + 13J_{d-d} + 5J_{s-d} + 25U'_{d-d},
\end{eqnarray}

\begin{eqnarray}
\label{a3}
E_{S=1/2}^{3d^94s^0} = 2\varepsilon_{A_1} + 3\varepsilon_{E_1} + 4\varepsilon_{E_2} \quad \quad \quad \quad \quad \nonumber \\
+ 4U_{d-d} + 16J_{d-d} + 32U'_{d-d},
\end{eqnarray}

\begin{eqnarray}
\label{a4}
E_{S=1/2}^{3d^74s^2} = 2\varepsilon_{A_1} + 2\varepsilon_{E_1} + 3\varepsilon_{E_2} + \varepsilon_{4s}  \quad \quad \quad \quad \nonumber \\
+ 2U_{d-d} + 9J_{d-d} + 7J_{s-d} + 19U'_{d-d},
\end{eqnarray}

\begin{eqnarray}
\label{a5}
E_{S=1/2}^{3d^84s^1} = \varepsilon_{A_1} + 4\varepsilon_{E_1} + 3\varepsilon_{E_2} + \varepsilon_{4s}  \quad \quad \quad \quad \nonumber \\
+ 3U_{d-d} + 13J_{d-d} + 3J_{s-d} + 25U'_{d-d}.
\end{eqnarray}

Substracting Eq.~(\ref{a1}) from Eq.~(\ref{a4}), and Eq.(\ref{a2}) from Eq.~(\ref{a5}), an expression for the exchange parameters can be readily obtained:
\begin{eqnarray}
\label{a6}
J_{d-d} = \frac{1}{2} \left( E_{S=3/2}^{3d^74s^2} - E_{S=1/2}^{3d^74s^2} \right),
\end{eqnarray}

\begin{eqnarray}
\label{a7}
J_{s-d} = \frac{1}{2} \left( E_{S=3/2}^{3d^84s^1} - E_{S=1/2}^{3d^84s^1} \right).
\end{eqnarray}

Summing up Eqs.~(\ref{a1}) and (\ref{a3}), subtracting Eq.~(\ref{a2}) twice, and exploiting the rotation invariance condition ($U'_{d-d} = U_{d-d} - 2J_{d-d}$), one obtain an expression for
the intrasite Coulomb repulsion parameter:
\begin{eqnarray}
\label{a8}
U_{d-d} = E_{S=3/2}^{3d^74s^2} +  E_{S=1/2}^{3d^94s^0} - 2E_{S=3/2}^{3d^84s^1} \quad \quad \quad \nonumber \\
- 2\Delta_{A_1-E_1}  - \Delta_{E_2-E_1} + J_{d-d} + 3J_{s-d},
\end{eqnarray}
where $\Delta_{i-j} = \varepsilon_i - \varepsilon_j$ are crystal-field splitting parameters.
Therefore, assuming that the rotation invariance condition is valid for the system under investigation, the electron-electron interaction parameters can be extracted from the total energies of 
different states up to the crystal-field splitting parameters.


\begin{thebibliography}{99}
\bibitem{Ahlrichs2}
The Ahlrichs (2$df$) polarization functions were obtained from the TurboMole
basis set library under ftp.chemie.uni-karlsruhe.de/pub/basen.
\bibitem{delta-calc}
Since the $\pi$ orbitals of graphene are approximated in this work by two degenerate $\pi_b$ orbitals of coronene, the repulsion between the 4$s$ electrons of Co and $\pi$ electrons of the surface at a 
distance $d$ can be quantified by the splitting between the corresponding HF energy levels, calculated relative to the splitting in the noninteracting system, i.e., as
$\delta_{s-\pi}^{\mathrm{HF}}(d)=\Delta_{4s-\pi}^{\mathrm{HF}}(d) - \Delta_{4s-\pi}^{\mathrm{HF}}(\infty)$, where $\Delta_{4s-\pi}^{\mathrm{HF}}=\varepsilon^{\mathrm{HF}}_{4s}-\varepsilon^{\mathrm{HF}}_{\pi_b}$.
\end{thebibliography}
\end{document}